\documentclass{emulateapj}

\def\ts     {\thinspace}
\def\kms    {\ts km\ts s$^{-1}$}
\def\etal   {{\rm et\ts al.}}
\def\msol   {M$_{\odot}$}

\def\ci     {C\ts {\scriptsize I}}

\def\aco    {{\rm CO}($J$=1$\to$0)}

\def\cco    {{\rm CO}($J$=3$\to$2)}

\def\gco    {{\rm CO}($J$=7$\to$6)}

\def\ahcn    {{\rm HCN}($J$=1$\to$0)}

\def\dhcn    {{\rm HCN}($J$=4$\to$3)}

\def\ahco    {{\rm HCO$^+$}($J$=1$\to$0)}

\def\acn    {{\rm CN}($N$=1$\to$0)}
\def\ccn    {{\rm CN}($N$=3$\to$2)}
\def\dcn    {{\rm CN}($N$=4$\to$3)}

\slugcomment{draft version \today, accepted for publication in the
  Astrophysical Journal}

\shorttitle{CN($N$=3--2) in the Cloverleaf}
\shortauthors{Riechers et al.}

\begin{document}

\title{Detection of Emission from the CN Radical in the Cloverleaf Quasar at $z$=2.56}

\author{Dominik A. Riechers\altaffilmark{1}, Fabian
  Walter\altaffilmark{1}, Pierre Cox\altaffilmark{2}, Christopher
  L. Carilli\altaffilmark{3}, \\  Axel Weiss\altaffilmark{4}, 
Frank Bertoldi\altaffilmark{5}, and Roberto Neri\altaffilmark{2}
}

\altaffiltext{1}{Max-Planck-Institut f\"ur Astronomie, K\"onigstuhl 17, 
Heidelberg, D-69117, Germany}

\altaffiltext{2}{Institut de RadioAstronomie Millim\'etrique, 300 Rue
  de la Piscine, Domaine Universitaire, 38406 Saint Martin d'H\'eres,
  France}

\altaffiltext{3}{National Radio Astronomy Observatory, PO Box O,
  Socorro, NM 87801, USA}

\altaffiltext{4}{Max-Planck-Institut f\"ur Radioastronomie, Auf dem
  H\"ugel 69, Bonn, D-53121, Germany}

\altaffiltext{5}{Argelander-Institut f\"ur Astronomie, Universit\"at
  Bonn, Auf dem H\"ugel 71, Bonn, D-53121, Germany}

\email{riechers@mpia.de}

\begin{abstract}
  We report the detection of \ccn\ emission towards the Cloverleaf
  quasar ($z=2.56$) based on observations with the IRAM Plateau de
  Bure Interferometer.  This is the first clear detection of emission
  from this radical at high redshift.  CN emission is a tracer of
  dense molecular hydrogen gas ($n({\rm H_2}) > 10^4\,$cm$^{-3}$)
  within star--forming molecular clouds, in particular in regions
  where the clouds are affected by UV radiation.  The HCN/CN intensity
  ratio can be used as a diagnostic for the relative importance of
  photodissociation regions (PDRs) in a source, and as a sensitive
  probe of optical depth, the radiation field, and photochemical
  processes.  We derive a lensing--corrected \ccn\ line luminosity of
  $L'_{\rm CN(3-2)} = (4.5 \pm 0.5) \times 10^{9}\,$K \kms pc$^2$.
  The ratio between CN luminosity and far--infrared luminosity falls
  within the scatter of the same relationship found for low--$z$
  (ultra--) luminous infrared galaxies.  Combining our new results
  with \cco\ and \ahcn\ measurements from the literature and assuming
  thermal excitation for all transitions, we find a CO/CN luminosity
  ratio of 9.3 $\pm$ 1.9 and a HCN/CN luminosity ratio of 0.95 $\pm$
  0.15.  However, we find that the \ccn\ line is likely only
  subthermally excited, implying that those ratios may only provide
  upper limits for the intrinsic 1$\to$0 line luminosity ratios.  We
  conclude that, in combination with other molecular gas tracers like
  CO, HCN, and HCO$^+$, CN is an important probe of the physical
  conditions and chemical composition of dense molecular environments
  at high redshift.
\end{abstract}


\section{Introduction}

Investigations of the dense molecular interstellar medium (ISM) in
high--redshift galaxies are of fundamental importance to further our
understanding of the early phases of galaxy formation and evolution,
as it harbors the environments in which the actual star formation is
believed to occur. Recent studies of the molecular gas phase in
high--$z$ galaxies using CO emission lines have revealed large
molecular gas reservoirs with masses in excess of 10$^{10}$\,\msol\
(see review by Solomon \& Vanden Bout \citeyear{sv05}, and references
therein).

However, while the relative brightness of CO emission lines renders
this molecule the most common tracer of the requisite material for
star formation, the low densities of only $n_{\rm H_2} \sim
10^2-10^3\,$cm$^{-3}$ required to excite its lower--$J$ transitions
(due to its low dipole moment of only $\mu_{\rm D}^{\rm CO} = 0.11$)
imply that CO is not a specific tracer of dense molecular cloud cores,
i.e., the regions where stars are actively formed.

In contrast, recent studies of nearby actively star--forming galaxies
have shown that the high dipole molecule HCN ($\mu_{\rm D}^{\rm HCN} =
2.98$) is a far better tracer of such dense molecular cores (e.g.\ Gao
\& Solomon \citeyear{gao04a}, \citeyear{gao04b}). The critical density
of $n_{\rm H_2} \sim 10^5-10^6$\,cm$^{-3}$ to collisionally thermalize
its lower--$J$ transitions is much higher than that of CO, and of the
same order as the densities found in bright star--forming regions in
the Galaxy (e.g.~the Orion Bar, Hogerheijde et al.\ \citeyear{hog95}).
A main result in this context is the finding that the HCN luminosity
correlates well with the far--infrared (FIR) luminosities (which is
commonly used to estimate star--formation rates at high redshift) over
7--8 orders of magnitude, from Galactic dense cores to the highest
redshift quasars (Wu et al.\ \citeyear{wu05}).

However, in nearby luminous and ultra--luminous infrared galaxies
(LIRGs/ULIRGs), it has been found that systems with similar HCN/CO and
HCN/FIR luminosity ratios may have quite different dense gas
properties regarding their chemical composition, and their gas
excitation (e.g., Aalto et al.\ \citeyear{aal02}). To better
understand the physical and chemical state of the dense molecular gas
phase which directly relates to star formation, it has proven
essential to study additional bright tracers of dense gas with
properties different from HCN, such as the cyanide radical (CN). Due
to its lower dipole moment relative to HCN ($\mu_{\rm D}^{\rm CN} =
1.45$), its critical density is lower by about a factor of 5.
Observations of CN emission toward the Orion A molecular cloud complex
have shown that CN filaments trace the dense interfaces between the
molecular cloud and the major ionization fronts (Rodr\'iguez--Franco
et al.\ \citeyear{rod98}).  It has also been found that the [CN]/[HCN]
abundance ratio is greatly enhanced in the central region of the
starburst galaxy M82, being as high as $\sim$5 across the entire
nucleus (Fuente et al.\ \citeyear{fue05}).  These observations
indicate that CN is a good tracer of gas layers which are affected by
photochemistry, since this molecule appears to be predominantly found
in regions exposed to ionizing stellar UV radiation. It has also been
found that, due to the rapid destruction of other dense gas tracers
like HCN, the abundance of CN tends to be enhanced in areas where the
UV radiation field is only partly attenuated, such as in zones close
to the surface of photodissociation regions (PDRs, Fuente et al.\
\citeyear{fue93}; Sternberg \& Dalgarno \citeyear{sd95}; Jansen et
al.\ \citeyear{jan95}).  The CN/HCN intensity ratio can thus be used
as a diagnostic for the relative importance of PDRs in a source, and a
sensitive probe of optical depth, the radiation field, and
photochemical processes (e.g., Boger \& Sternberg \citeyear{bs05}).
CN emission may thus in a sense be a more specific tracer for star
formation than, e.g., HCN, which only traces regions of dense gas in
general. In addition, theoretical studies of the chemical composition
of molecular gas for different ionization parameters suggest that the
relative abundance of CN may also be enhanced in X--ray dominated
regions (XDRs), such as AGN environments (Lepp \& Dalgarno
\citeyear{lep96}). This exemplifies why it is desirable to search for
CN emission, which is part of our current effort to study molecular
tracers other than CO out to high redshifts. Such an investigation is
imperative to obtain more meaningful constraints on the physical
properties and chemical composition of the dense molecular ISM in
distant galaxies.

In this paper, we report the first high--$z$ detection of \ccn\
emission, which was observed towards the Cloverleaf quasar ($z=2.56$)
with the IRAM Plateau de Bure Interferometer (PdBI)\footnote{IRAM is
  supported by INSU/CNRS (France), MPG (Germany), and IGN (Spain).}.
Due to its strong gravitational magnification (magnification factor
$\mu_{\rm L} = 11$, Venturini \& Solomon \citeyear{ven03}), the
Cloverleaf is the brightest CO source at high redshift (e.g.,
Barvainis et al.\ \citeyear{bar94}), and one of the most prolific
sources of molecular lines beyond $z$=2. It was the first $z>2$ source
to be detected in HCN (Solomon \etal\ \citeyear{sol03}) and HCO$^+$
(Riechers et al.\ \citeyear{rie06a}) emission. It was also detected in
both \ci\ fine structure lines (Barvainis et al.\ \citeyear{bar97};
Wei\ss\ \etal\ \citeyear{wei03}, \citeyear{wei05}). We use a standard
concordance cosmology throughout, with $H_0 = 71\,$\kms\,Mpc$^{-1}$,
$\Omega_{\rm M} =0.27$, and $\Omega_{\Lambda} = 0.73$ (Spergel \etal\
\citeyear{spe03}, \citeyear{spe06}).

\section{Observations}

We observed the \ccn\ transition line ($\nu_{\rm rest} =
339.4467770-340.2791661\,$GHz for the different fine structure [fs]
and hyperfine structure [hfs] transitions of the $v$=0 vibrational
state) towards H1413+117 (the Cloverleaf quasar) using the PdBI in D
configuration between 2006 July 26 and September 03.  At the target
redshift of $z$=2.55784 (Wei\ss\ \etal\ \citeyear{wei03}), the line is
shifted to $\sim$95.6\,GHz (3.14\,mm).  The total integration time
amounts to 17.5\,hr using 5 antennas, resulting in 5.8\,hr equivalent
on--source time with 6 antennas after discarding unusable visibility
data. The nearby sources 1354+195 and 1502+106 (distance to the
Cloverleaf: $9.0^\circ$ and $12.0^\circ$) were observed every 20
minutes for pointing, secondary amplitude and phase calibrations.  For
primary flux calibration, several nearby calibrators (MWC\,349,
CRL\,618, 3C\,273, 3C\,345, and NRAO\,150) were observed during all
runs.

The correlator was tuned to a frequency of 95.603\,GHz, which
corresponds to the central position between the brightest hfs
transitions\footnote{These components are actually blends of the
  [$N$=3$\to$2, $J$=$\frac{5}{2}$$\to$$\frac{3}{2}$,
  ($F$=$\frac{7}{2}$$\to$$\frac{5}{2}$,
  $\frac{5}{2}$$\to$$\frac{3}{2}$, and
  $\frac{3}{2}$$\to$$\frac{1}{2}$)] hfs components at $\nu_{\rm rest}
  = 340.0315440-340.0354080\,$GHz and [$N$=3$\to$2,
  $J$=$\frac{7}{2}$$\to$$\frac{5}{2}$,
  ($F$=$\frac{9}{2}$$\to$$\frac{7}{2}$,
  $\frac{7}{2}$$\to$$\frac{5}{2}$, and
  $\frac{5}{2}$$\to$$\frac{3}{2}$)] hfs components at $\nu_{\rm rest}
  = 340.2477700-340.2485764\,$GHz (components 2 and 3 in
  Fig.~\ref{f1}).}  of the \ccn\ line at 95.6332 and 95.5736\,GHz. The
total bandwidth of 580\,MHz ($\sim$1800\,\kms ) used for the
observations is large enough to cover all fs and hfs transitions of
the \ccn\ line. It also covers enough channels that are free of line
emission to constrain the 3\,mm continuum emission of the Cloverleaf.

For data reduction and analysis, the IRAM GILDAS package was used.
All data were mapped using the CLEAN algorithm and 'natural' weighting
without applying a further taper; this results in a synthesized beam
of 5.3\,$''$$\times$4.9\,$''$ ($\sim$42\,kpc at $z$ = 2.56).  The
final rms in the combined map is 0.25\,mJy beam$^{-1}$ for a 190\,MHz
(corresponding to 596\,\kms) channel, 0.4\,mJy beam$^{-1}$ for a
80\,MHz (252\,\kms) channel, and 0.8\,mJy beam$^{-1}$ for a 20\,MHz
(63\,\kms) channel.

\section{Theoretical Considerations}

\begin{figure}[t]
\epsscale{1.2}
\plotone{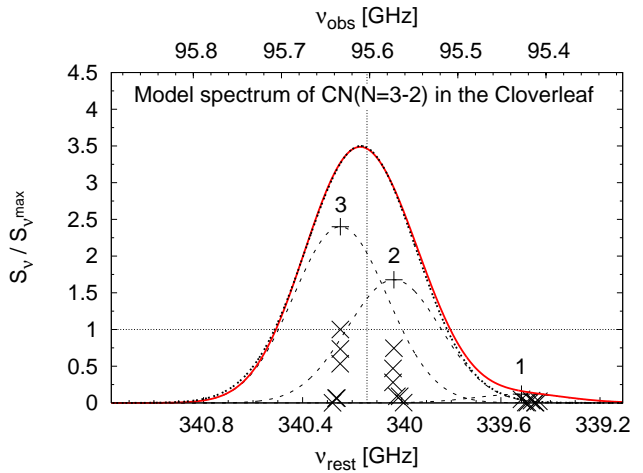}
\caption{Model spectrum of the \ccn\ emission in the Cloverleaf. The
  horizontal axes indicate the rest and observed frequencies. The full
  range of the PdBI bandpass is shown. The vertical axis indicates the
  predicted intensity, normalized to the ($N$=3$\to$2,
  $J$=$\frac{7}{2}$$\to$$\frac{5}{2}$,
  $F$=$\frac{9}{2}$$\to$$\frac{7}{2}$) component. The relative
  intensities are computed for the LTE case. The crosses indicate the
  hyperfine structure components.  The plus signs indicate the summed
  intensities of the components in three different frequency bins.
  The dashed lines are Gaussian fits to the summed intensities,
  assuming the width of the \cco\ line (Wei\ss\ et al.\
  \citeyear{wei03}) for the subcomponents (labeled '1'--'3'). The
  solid line is a sum of all three components, and indicates the
  model--predicted \ccn\ line shape.  The thick dotted line is a
  single Gaussian (used to fit the observations), fitted to the model
  line.  The dotted vertical line indicates the tuning frequency,
  corresponding to zero velocity in Fig.~\ref{f4}. \label{f1}}
\end{figure}

Due to its fine stucture and hyperfine structure splitting, the \ccn\
emission line is distributed over 19 lines\footnote{We assume that CN
  is in its ground electronic state (${}^2\Sigma$), and that the spins
  couple according to Hund's case (b) coupling scheme: $\vec{N} +
  \vec{S} = \vec{J}$, (fs coupling) and $\vec{J} + \vec{I_{\rm C}} +
  \vec{I_{\rm N}} = \vec{F}$ (hfs coupling). Here, $\vec{N}$ is the
  rotational angular momentum vector, $\vec{S}$ is the electronic
  spin, and $\vec{I}$ is a nuclear spin.}, in 3 main components
separated by more than 200\,MHz from each other in the rest frame.
This separation is of the same order as the kinematical broadening of
the CO lines in the Cloverleaf (Wei\ss\ et al. \citeyear{wei03}),
causing the hfs components to be blended.  To analyze the intrinsic
line shape of the \ccn\ transition in the Cloverleaf, we calculated a
synthetic line profile, assuming optically thin emission in Local
Thermodynamic Equilibrium (LTE) to derive the relative intensities of
the hfs components. The relative intensities of the components were
computed using the laboratory data from Skatrud et al.\
(\citeyear{ska83}), and approximation (4) of equation (1) of Pickett
et al.\ (\citeyear{pic98}).  Assuming that the 3 main \ccn\ components
are kinematically broadened in the same way as the CO lines (416 $\pm$
6\,\kms\ FWHM; see Wei\ss\ et al. \citeyear{wei03}), we obtain the
synthetic line profile displayed in Fig.~\ref{f1} (solid line).  Under
the given assumptions, the contribution from all hfs lines of
component 1 to the total intensity are negligible, and not detectable
at the given signal--to--noise (see below).  A single Gaussian fit
(dotted line, 484\,\kms\ FWHM, or 116\% of the CO lines) agrees with
the more detailed model profile (solid line) within a few percent in
its peak position and integral, i.e., well within the observational
errors. Note however that the width and peak position of this Gaussian
depends on the relative intensities of the hfs components, which in
turn depend on the above assumptions, in particular the optical depth.
In the optically thin LTE case, components 2 and 3 have a peak
strength ratio of 1:1.4.  As an example, if they had a ratio of unity,
the fitted Gaussian to the line profile would have a FWHM of
603\,\kms, or 145\% of the CO lines.  We thus conclude that fitting a
Gaussian to the observed line profile contains all relevant
information, while minimizing the number of free fit parameters, and
thus is preferred over a more complex fitting procedure to describe
and analyze the \ccn\ profile at the given signal--to--noise ratio.

\section{Results}

\subsection{CN Maps and Spectrum}

\begin{figure}
\epsscale{1.15}
\plotone{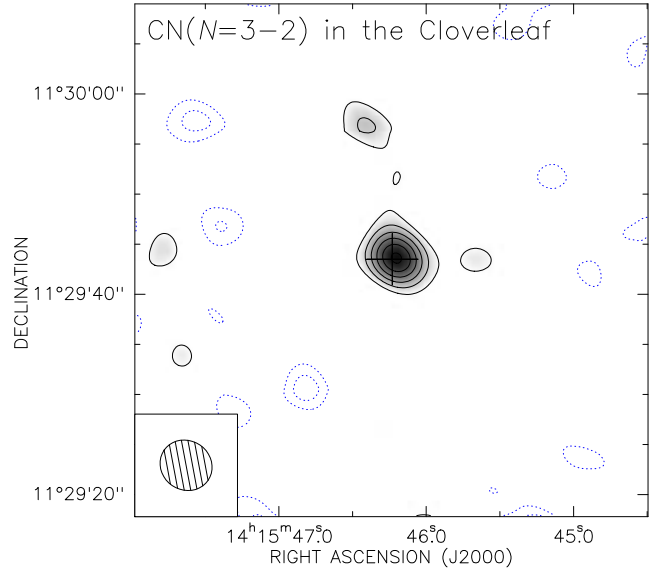}
\caption{Velocity--integrated PdBI map of \ccn\ emission towards the
  Cloverleaf quasar.
  At a resolution of 5.3\,$''$$\times$4.9\,$''$
  (as indicated in the bottom left corner), the source is 
  unresolved. The cross indicates the geometrical center of the CO
  emission in the Cloverleaf (Alloin et al.\ \citeyear{all97}, see
  text).  Contours are shown at (-3, -2, 2, 3, 4, 5, 6,
  7)$\times\sigma$ (1$\sigma$ = 0.25\,mJy beam$^{-1}$). 
  \label{f2}}
\end{figure}

\begin{figure*}
\epsscale{1.15}
\plotone{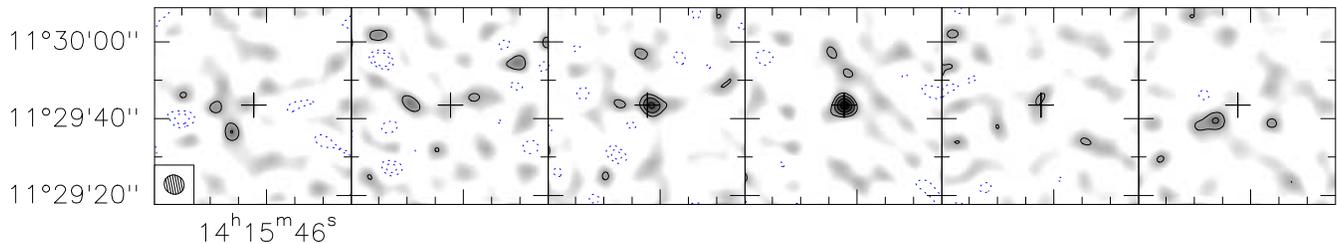}
\caption{Channel maps of the \ccn\ emission (same region is shown as
  in Fig.~\ref{f2}). One channel width is 80\,MHz, or
  252\,km\,s$^{-1}$ (at 95.4355, 95.5155, 95.5955, 95.6755, 95.7555,
  and 95.8355\,GHz; frequencies increase with channel number).
  Contours are shown at (-3, -2, 2, 3, 4, 5)$\times \sigma$ (1$\sigma$
  = 0.4\,mJy\,beam$^{-1}$). The beam size (5.3\,$''$$\times$4.9\,$''$)
  is shown in the bottom left corner; the cross indicates the same
  position as in Fig.~\ref{f2}.
  \label{f3}}
\end{figure*}

We have detected emission from the \ccn\ transition line towards the
Cloverleaf quasar ($z=2.56$).  The velocity--integrated \ccn\ line map
is shown in Fig.~\ref{f2}.  The cross indicates the geometrical
center position of the resolved \gco\ map at $\alpha=14^{\rm h}15^{\rm
  m}46^{\rm s}.233$, $\delta=+11^\circ29'43''.50$ (Alloin \etal\
\citeyear{all97}).  The line emission is clearly detected at 7$\sigma$
over a range of 596\,\kms\ (190\,MHz); the source appears unresolved.
In Fig.~\ref{f3}, six channel maps (252\,\kms, or 80\,MHz each) of the
\ccn\ line emission are shown.  At an rms of 0.4\,mJy\,beam$^{-1}$,
the line is detected at 4 and 5$\sigma$ in the central channels, and
the decline of the line intensity towards the line wings is clearly
visible toward the outer channels, as expected.

\begin{figure}[b]
\epsscale{1.15}
\plotone{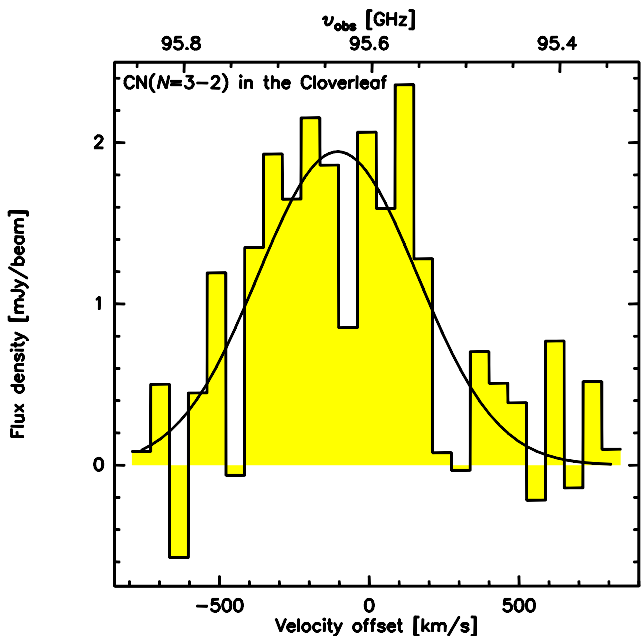}
\caption{ Spectrum of the \ccn\ emission at a resolution of 63\,\kms\
  (20\,MHz).  The velocity scale is relative to the tuning frequency
  of 95.603\,GHz.  The rms per velocity bin is 0.8\,mJy. The solid
  line shows a Gaussian fit to the data. \label{f4}}
\end{figure}

Fig.~\ref{f4} shows the spectrum of the \ccn\ emission at a
resolution of 63\,\kms\ (20\,MHz). Zero velocity corresponds to the
tuning frequency of 95.603\,GHz.  The solid line shows a Gaussian fit
to the spectrum. From the fit, we derive a line peak flux density of
1.94 $\pm$ 0.24\,mJy beam$^{-1}$, and a line FWHM of 666 $\pm$
97\,\kms.  The fit provides an upper limit for the continuum emission
at the line frequency of $<$0.25\,mJy.  From the channels assumed to
be free of line emission, we derive a formal 3$\sigma$ upper limit to
the continuum peak flux density of 0.7\,mJy beam$^{-1}$.  Note that
the model fit to the dust SED of the Cloverleaf by Wei\ss\ et al.\
(\citeyear{wei03}) would suggest a continuum flux of $\sim$0.3\,mJy at
the \ccn\ line frequency, which agrees within the errors with the
above estimates.  As the continuum however is not detected, we do not
subtract a continuum component from the observed spectrum.  This leads
to an integrated \ccn\ line flux of 1.37 $\pm$ 0.17\,Jy \kms.  The
velocity offset of the Gaussian peak relative to the tuning frequency
is -105 $\pm$ 38\,\kms.

The FWHM of the Gaussian fit to the \ccn\ emission line suggests that
its components 2 and 3 have similar peak strengths (different from
1:1.4 as predicted for the optically thin LTE case; Fig.~\ref{f1}),
which may indicate that the emission is optically thick.  Within the
limited signal--to--noise of the spectrum, the shape of the emission
line remains compatible with a two--component structure. It may thus
be possible that the two brightest hfs complexes (components 2 and 3)
are detected individually, but observations at higher
signal--to--noise are needed to confirm this result. In any case, a
double Gaussian fit to the line profile gives the same integrated line
flux as the single Gaussian fit within the errors.

\subsection{Line Luminosities}

From our observations, we derive a \ccn\ line luminosity of $L'_{\rm
  CN(3-2)} = (4.5 \pm 0.5) \times 10^9\,$K\,\kms\,pc$^2$ (corrected
for gravitational magnification, $\mu_{\rm L} = 11$; see Table 1 for
details). This corresponds to 11\% of $L'_{\rm CO(3-2)}$ (Wei\ss\ et
al.\ \citeyear{wei03}), 105\% of $L'_{\rm HCN(1-0)}$ (Solomon et al.\
\citeyear{sol03}), and 129\% of $L'_{\rm HCO^+(1-0)}$ (Riechers et
al.\ \citeyear{rie06a}).  Assuming that CN is thermally excited up to
the $N$=3$\to$2 transition (i.e., $L'_{\rm CN(3-2)} = L'_{\rm
  CN(1-0)}$) and that optical depth effects can be neglected, this
would mean that CN emission in the Cloverleaf is slightly brighter
than that from the other high density probes HCN and HCO$^+$. This is
remarkable, as HCN and HCO$^+$ have higher critical densities than CN
in the ground--state transition, but the critical density of \ccn\ is
higher than that of \ahcn\ and \ahco.  This may indicate a higher
relative filling factor of CN relative to HCN and HCO$^+$, or even a
relatively high chemical abundance of CN.  However, with the
observations existing at present, it is not possible to disentangle
excitation effects from chemical effects.

In addition, observations of \dhcn\ towards the Cloverleaf quasar have
shown that $L'_{\rm HCN(4-3)}$/$L'_{\rm HCN(1-0)} \leq 0.34$ (Solomon
et al.\ \citeyear{sol03}, caption of their Tab.~1, and M.~Gu\'elin
2007, priv.~comm.). This ratio is significantly lower than 1, which
implies that the 4$\to$3 transition of HCN is sub--thermally excited.
As the \ccn\ transition has a somewhat lower but comparable critical
density relative to \dhcn, it is likely that the \ccn\ transition is
also sub--thermally excited, and the `intrinsic',
thermalization--corrected $L'_{\rm CN(1-0)}$ thus even higher.  This
would be consistent the finding that CN is clearly subthermally
excited in nearby LIRGs and ULIRGs (Aalto et al.\ \citeyear{aal02}).
This would also imply that CN is a {\em brighter} tracer of dense gas
than HCN and HCO$^+$ in this high redshift quasar.

\begin{deluxetable}{ r c c c }
\tabletypesize{\scriptsize}
\tablecaption{Molecular line luminosities in the Cloverleaf. \label{tab-1}}
\tablehead{
& $S_{\nu}$ & $L'$ & Ref. \\
& [mJy] & [10$^9$\,K\,\kms\,pc$^2$] & }
\startdata
\ccn\ & 1.94 $\pm$ 0.24 & 4.5 $\pm$ 0.5 & 1 \\
\ahcn\ & 0.24 $\pm$ 0.04 & 4.3 $\pm$ 0.5 & 2 \\
\ahco\ & 0.19 $\pm$ 0.03 & 3.5 $\pm$ 0.3 & 3 \\
\cco\ & 30 $\pm$ 1.7 & 42 $\pm$ 7 & 4 \\ 
\vspace{-2mm}
\enddata 
\tablerefs{${}$ [1] This work, [2] Solomon \etal\ (\citeyear{sol03}),
  [3] Riechers et al.\ (\citeyear{rie06a}), [4] Wei\ss\ \etal\
  (\citeyear{wei03}).}
\tablecomments{${}$Luminosities are corrected for gravitational
  magnification.  }
\end{deluxetable}


\section{Discussion}

\begin{figure*}
\epsscale{1.15}
\plotone{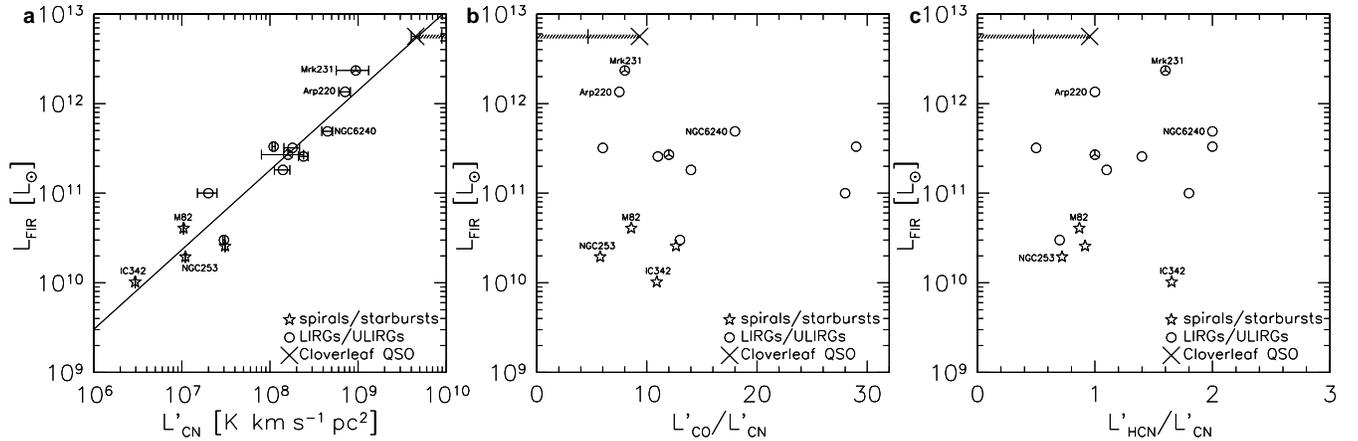}
\caption{CN luminosity relations for a sample of local
spiral/starburst galaxies, low--$z$ IR--luminous galaxies, and the
Cloverleaf.  The Cloverleaf luminosities are corrected for
gravitational lensing ($\mu_{\rm L} = 11$). The two symbols with star
insets denote tentative CN detections.  The solid line is a least
squares fit to all data except the Cloverleaf.  The error bars
indicate the statistical errors of the line luminosity measurements.
The excitation of the \ccn\ transition in the Cloverleaf may be
sub--thermal, which would affect the (extrapolated) \acn\ line
luminosity as indicated by the horizontal, shaded regions. As an
example, the vertical bars on the shaded regions show a case where
$L'_{\rm CN(3-2)} = 0.5 \times L'_{\rm CN(1-0)}$ in the Cloverleaf.
See text for more details. \label{f5}}
\end{figure*}

In the following, we discuss various relationships between the
emission observed in CN and other molecules (CO and HCN) and the
far--IR continuum for a sample of local spiral/starburst galaxies
(Henkel et al.\ \citeyear{hen88}, \citeyear{hen98}, 2007, priv.~comm.;
Wang et al.\ \citeyear{wan04}; additional CO/HCN/FIR data from Eckart
et al.\ \citeyear{eck90}; Nguyen-Q-Rieu et al.\ \citeyear{ngu92};
Mauersberger et al.\ \citeyear{mau96}; Mao et al.\ \citeyear{mao2000};
Sanders et al.\ \citeyear{san03}), low--$z$ (U)LIRGs (Aalto et al.\
\citeyear{aal02}), and the Cloverleaf (this work; Solomon et al.\
\citeyear{sol03}; Wei\ss\ et al.\ \citeyear{wei03}) as shown in
Fig.~\ref{f5}. Note that, due to the weak signal--to--noise ratio of
1$\sigma$, we do not include the \dcn\ observations of APM\,08279+5255
(Guelin et al.\ \citeyear{gue06}, their Tab.~1) in this discussion.

As \aco\ is not yet detected toward the Cloverleaf, we assume in the
following that CO is thermalized up to the 3$\to$2 transition, i.e.,
$L'_{\rm CO(1-0)} = L'_{\rm CO(3-2)}$ (see Riechers et al.\
\citeyear{rie06b} for justification).  Both effects of thermal and
sub--thermal excitation of the \ccn\ emission line in the Cloverleaf
are discussed.  CO, CN, and HCN line intensities for all low--$z$
galaxies were measured directly in the ground--state transitions, all
quoted luminosities thus are 1$\to$0 luminosities.  We do not discuss
effects of differential lensing\footnote{Note that Chartas et al.\
  (\citeyear{cha07}) recently reported a likely microlensing event
  toward the Cloverleaf; however, this event took place close to 2000
  April, so 3--7\,years before all of the molecular gas observations
  of this source discussed in this paper (obtained between 2003 March
  and 2006 September).}, as models indicate similar sizes for
molecular and dust emission in the Cloverleaf (Solomon et al.\
\citeyear{sol03}).

{\em Figure 5a}: Wu et al.\ (\citeyear{wu05}) find a linear
correlation between $L'_{\rm HCN}$ and $L_{\rm FIR}$ over 7--8 orders
of magnitude in luminosity, which is believed to indicate that the
dense gas tracer HCN is also a good estimator for star formation rates
out to high $z$. If CN were to be a good tracer of dense, actively
star--forming gas, a similar trend may be expected between $L'_{\rm
  CN}$ and $L_{\rm FIR}$. Figure~\ref{f5}a shows that $L'_{\rm CN}$
correlates closely with $L_{\rm FIR}$; a linear least squares fit to
all galaxies (excluding the Cloverleaf) yields log$(L_{\rm FIR}) =
(0.89 \pm 0.09) \times {\rm log} (L'_{\rm CN})+(4.2 \pm 0.7)$.  This
slope may suggest a trend of decreasing CN/FIR luminosity ratio
towards higher luminosities (see also Aalto et al.\ \citeyear{aal02}).
It however also is consistent with unity within the statistical
uncertainties (which are quite large due to the limited sample size)
and systematical errors (e.g., different beam sizes for CN line and IR
continuum observations), and thus with CN being a valuable tracer of
star formation.  The Cloverleaf agrees remarkably well with the
relation defined by the low--$z$ galaxies (even if the \ccn\
transition is sub--thermally excited, as indicated by the horizontal
bar on the shaded region), extending the observed trend to higher
luminosities and out to high redshift.

{\em Figure 5b}: Gao \& Solomon (\citeyear{gao04b}) observe a rise in
dense gas fraction with $L_{\rm FIR}$.  The $L'_{\rm CO}$/$L'_{\rm
  CN}$ ratio may be considered (the inverse of) a tracer of the dense
gas fraction. A significant spread in the $L'_{\rm CO}$/$L'_{\rm CN}$
ratio is found for the sample shown in Fig.~\ref{f5}b, ranging from a
few to almost a factor of 30. There is no obvious trend of $L'_{\rm
  CO}$/$L'_{\rm CN}$ with $L_{\rm FIR}$; however, the sample size is
too small and heterogenous to come to a definite conclusion.  Note
that, due to the likely sub--thermal excitation of the \ccn\
transition (as indicated by the shaded region), the $L'_{\rm
  CO}$/$L'_{\rm CN}$ ratio may be considered an upper limit of the
`intrinsic', thermalization--corrected ratio of the 1$\to$0
transitions for the Cloverleaf.

{\em Figure 5c}: Assuming that CN traces the UV field (which falls off
rapidly with growing distance from star--forming regions, or an active
galactic nucleus), while HCN traces dense gas in general, it would be
expected that the HCN/CN ratio stays constant or decreases with
increasing $L_{\rm FIR}$. In this scenario, a decrease in the HCN/CN
ratio would correspond to an increase in the filling factor of
UV--illuminated clouds with $L_{\rm FIR}$. This would however also
cause a rise in $L'_{\rm CN}$/$L_{\rm FIR}$, which is not observed
(Fig.~\ref{f5}a). If differences in the relative chemical abundances
of CN and HCN and optical depth/thermalization effects do not play a
major role, one may thus expect $L'_{\rm HCN}$/$L'_{\rm CN}$ to stay
fairly constant with $L_{\rm FIR}$.  Based on the observations of
their subsample, Aalto et al.\ (\citeyear{aal02}) suggest that the
HCN/CN intensity ratio may increase slightly with $L_{\rm FIR}$.  Such
a trend however is not seen in the larger sample displayed in
Fig.~\ref{f5}c (note that $L'_{\rm HCN}$/$L'_{\rm CN}$ for the
Cloverleaf may again be considered an upper limit due to possible
sub--thermal excitation of \ccn, as indicated by the shaded region).
Clearly, improved statistics and better models are required to address
this issue in more detail.

Chemical models suggest that CN is produced quite efficiently in the
presence of a strong UV field, which also leads to an enhanced
ionization rate (Boger \& Sternberg \citeyear{bs05}). The large
observed [CN]/[HCN] abundance ratio of $\sim$5 toward the molecular
disk of M82 may be indicative of a large dense PDR bathed in the
intense radiation field of the starburst environment (Fuente et al.\
\citeyear{fue05}), lending observational support to these models.
Other models suggest that the relative abundance of CN may also be
enhanced in regions with elevated X--ray ionization rates, such as AGN
environments (Lepp \& Dalgarno \citeyear{lep96}; Meijerink \& Spaans
\citeyear{mei05}).  Especially in the Cloverleaf, where the central
quasar is known to contribute a significant fraction to the heating of
the gas and dust (Wei\ss\ et al.\ \citeyear{wei03}; Solomon et al.\
\citeyear{sol03}), such a scenario would appear reasonable and may
explain the relatively high brightness of CN compared to other dense
gas tracers in the Cloverleaf. However, note that in Fig.~5c, the pure
starburst galaxies show the brightest CN emission relative to HCN, and
galaxies with a relatively strong AGN like Mrk\,231 and NGC\,6240 have
some of the highest $L'_{\rm HCN}$/$L'_{\rm CN}$ ratios. This may
indicate that the strength of the UV field in a starburst environment
has a significantly greater impact on the global gas--phase production
rate of CN in a galaxy than the strength of the X--ray field, lending
further support to the supposition that CN is a good star formation
tracer.  We thus conclude that the XDR scenario alone does not explain
the observed dense gas properties of the (U)LIRGs shown in Fig.~5
without any further assumptions.

Overall, the Cloverleaf follows the $L'_{\rm CN}$--$L_{\rm FIR}$
relation as defined by nearby galaxies of different types remarkably
well, and extends this relation to higher luminosities. This relation
now appears to hold over almost 3 orders of magnitude. If these
findings were to hold for other distant quasars, CN would be an
excellent probe of dense molecular environments out to high redshifts.
Also, the higher--order transitions of CN appear to be brighter than
those of other dense gas tracers like HCN. This may prove to be of
particular importance for future high--$z$ studies with the Atacama
Large Millimeter/submillimeter Array (ALMA), which will offer the
opportunity to probe to fainter galaxy populations in general, but
will also be restricted to the higher--order transitions of the most
common dense molecular gas tracers (typically 3$\to$2 and higher at
$z>2$).

\acknowledgments 
This research is based on observations carried out with the IRAM
Plateau de Bure Interferometer. IRAM is supported by INSU/CNRS
(France), MPG (Germany), and IGN (Spain). D.~R.\ acknowledges support
from the Deutsche Forschungsgemeinschaft (DFG) Priority Programme
1177.  C.~C.\ acknowledges support from the Max-Planck-Gesellschaft
and the Alexander von Humboldt-Stiftung through the
Max-Planck-Forschungspreis 2005. The authors would like to thank
Susanne Aalto for fruitful discussions on the subject matter.  We also
would like to thank the staff at IRAM, in particular Jan Martin
Winters, for their assistance in setting up and carrying out the
observations, as well as for providing a preliminary reduction of the
data.  The authors would like to thank the referee for useful comments
that helped to improve the manuscript.

\end{document}